\documentclass[11pt, oneside]{article}

\usepackage[T1]{fontenc}
\usepackage{graphicx}
\usepackage[margin=1in]{geometry} 
\usepackage{amsmath,amsfonts,amsthm,mathtools}
\usepackage{url} 
\usepackage{natbib} 
\usepackage{setspace} 
\usepackage{comment}
\def\ci{\perp\!\!\!\perp}

\newcommand{\indc}[1]{\mathbb{I}\left\{#1\right\}}

\title{A probabilistic formalization of contextual bias in forensic analysis: Evidence that examiner bias leads to systemic bias in the criminal justice system}

\author{
Maria Cuellar\\
Department of Criminology\\
University of Pennsylvania\\
Philadelphia, PA 19104 \\
\texttt{mcuellar@sas.upenn.edu} \\
\\
\and
Jacqueline Mauro \\
Data Science\\
Google, Inc.\\
Mountain View, CA 94043\\
\texttt{jacqueline.mauro@berkeley.edu} \\
\and
Amanda Luby \\
Department of Mathematics and Statistics\\
Swarthmore College\\
Swarthmore, PA 19081\\
\texttt{aluby1@swarthmore.edu } \\
}

\date{November 5, 2021}

\begin{document}

\maketitle
\onehalfspacing


\begin{abstract}
Although researchers have found evidence contextual bias in forensic science, the discussion of contextual bias is currently qualitative. We formalize years of empirical research and extend this research by showing quantitatively how biases can be propagated throughout the legal system, all the way up to the final determination of guilt in a criminal trial. We provide a probabilistic framework for describing how information is updated in a forensic analysis setting by using the ratio form of Bayes' rule. We analyze results from empirical studies using our framework and use simulations to demonstrate how bias can be compounded where experiments do not exist. We find that even minor biases in the earlier stages of forensic analysis lead to large, compounded biases in the final determination of guilt in a criminal trial.
\end{abstract}

\textbf{Key words}: Contextual bias, forensic science, Bayes' rule, decision-making, formalization.

\section{Introduction}

Contextual bias occurs in forensic science when well-intentioned forensic analysts are vulnerable to making errors when exposed to external information \citep{thompsonbias}. There is evidence of this type of bias in a broad variety of forensic disciplines, including fingerprints \citep{dror-mayfield}, bite marks \citep{osborne-bitemark}, and even the gold-standard of forensic disciplines, DNA \citep{dror-dnaevidence}. US government-led reports have also stated that analysts' judgments can be influenced by irrelevant information about the facts of a case \citep{nationalacademy}, and they urged that efforts be made to mitigate this bias by ensuring ``that examiners are not exposed to potentially biasing information'' (PCAST, 2016).

Contextual bias has been demonstrated even in very high profile cases. After the 2004 Madrid bombing, American attorney Brandon Mayfield was wrongfully accused of being involved in the attack after the FBI mistakenly concluded that Mayfield was the source of a fingerprint found at the crime scene. The FBI reviewed what could have led to this mistake and found that the factors to blame included over-reliance on the algorithm's proposed candidates, other analysts' opinions, the fact that it was a high-profile case, and Mayfield's religion, among others \citep{fbi}. In an experiment designed to measure contextual bias in fingerprint comparisons, analysts altered their conclusions after learning about contextual information related to the Mayfield case in four out of five cases \citep{dror-mayfield}.

It is unclear what is meant by this "bias" that led to the Mayfield error, and a lack of formalization can lead to disagreement about the definition of contextual bias and how to mitigate it. What do the biasing factors from the Mayfield case have in common? Which information is biasing, and should this information have been excluded in the material given to analysts? Do interactions between analysts lead to bias? How does the bias affect the final decision of guilt by a judge or jury? These are the types of questions that formalizing the forensic reasoning can help answer. 

Our contribution is to clarify the problem of contextual bias by turning the current qualitative discussion of bias into a rigorously defined and quantitative discussion. We provide a formal framework for how information is updated in cases of cognitive bias by using the ratio form of Bayes' rule. Bayes' rule is a useful tool for this because it mirrors the adversarial system of the courts, and for this reason it has been used elsewhere in forensics to motivate the use of likelihood ratio-based reporting \citep{Aitken2018, ommen}. We hope this formalization can serve as a tool for transparency, and is useful for both laboratories seeking to minimize bias in their analyses and to researchers developing algorithms to augment forensic decision-making tasks.

We also show how systemic bias in the courts is possible in the decision of rational, well-intentioned actors. Even if actors in the system did not discriminate with respect to race, ethnicity, income, sexual orientation, gender, or other protected characteristics themselves, they might still reach a biased conclusion. Thus, even without any biased forensic analysts, judges, or juries, there can still be bias in the final determination of guilt.

Section 2 presents the formal setup and notation, the definitions of the tasks undertaken at three different levels: the analyst, the laboratory, and the trier of fact. Section 3 describes biases in the analyst's conclusions due to several factors, including data, reference materials, and contextual information. Section 4 shows that bias can be propagated throughout the laboratory where the rest of the evidence is analyzed by analysts, including as a bias cascade or snowball. Section 5 describes how the trier of fact hears the analysts' conclusions to make a final decision about guilt.

\section{Formal Setup}\label{sec:setup}

\subsection{Notation}

In this section, we describe a hypothetical workflow of fingerprint analysis to develop a formal model of reasoning in forensic analysis. We use previously defined notation \citep{lundiyer, lindley}. Our approach is a generalization of forensic practice, since each forensic discipline and each laboratory have a specific workflow for analyzing evidence. We use fingerprint comparisons as a running example to describe our model, and this could then be applied to other forensic disciplines.

Suppose a latent print $y$ was left at a crime scene and a forensic team lifts this print. Police then identify a suspect and collect a fingerprint $x$ from that individual. In reality, police collect the ten prints from a suspect (or a set of suspects is found through a database search), and the analyst selects which print, if any, is most like the latent print. For this article, we assume that only one print is being compared. The latent print $y$ from the crime scene may have some sections that are blurry or distorted and some that are missing, while the exemplar print $x$ from the suspect is clean and complete, since if an exemplar print is defective, it may be taken again. The data contained in $y$ are a sample from a random variable $Y$ with the underlying distribution $F_y$, a relationship denoted by $Y \sim F_y$. As usual, capital letters denote random variables and lower-case letters denote observations of those random variables. Similarly, $X \sim F_x$.

At the forensic laboratory, the analyst is asked to analyze the fingerprints. Suppose $I$ denotes all the information available to the analyst, in addition to being shown $y$ and $x$. This set $I$ could include, for example, the fact that the supervisor mentioned that they believe the suspect committed the crime, information about the suspect's gender, age, or race, or the suspect's criminal history. It could also include information about the evidence itself, such as that the print was collected from a glass surface, or that the print was made up of a tacky material. 

Let $\mathcal{S} = \{S_0,S_1,\dots ,S_N\}$ denote the set of all the potential sources (individuals denoted by $0, \dots, N$) of the crime scene, latent print. Since the source of $y$ is unknown it is denoted by $S_q$, where $q$ is the questioned individual. The suspect is source $S_0$ of fingerprint $x$. Thus, the analyst's task is to determine whether $x$ and $y$ have a common source, $S_0$, an event denoted by $E_0$,
\begin{equation}
\label{}
E_0: S_0\text{ is the source of both $x$ and $y$ (i.e. $q=0$)},
\end{equation}
or whether they have a different source. That is, some $S_j, j\in\{ 1, \dots, N\}$ is the source of $y$ while $S_0$ is the source of $x$, an event denoted by the complement of $E_0$,
\begin{equation}
\label{}
E_0^\mathsf{c}: S_1\text{ or }S_2\text{ or \dots or } S_N \text{ is the source of $y$ (i.e.\ $q\in\{1, 2, \dots, N$\})}.
\end{equation}

\subsection{The analyst's task}

The analyst is responsible for determining whether two pieces of evidence were produced by the same source. In fingerprint analysis, this is usually reported in terms of categories: an \textit{identification} (i.e. a conclusion in favor of $E_0$), an \textit{exclusion} (a conclusion in favor of $E_0^c$), or an \textit{inconclusive} (a conclusion that favors neither $E_0$ or $E_0^c$). 

We formalize this task by using probabilities and the ratio form of Bayes' rule, as others have done before \citep{lundiyer, lindley, Aitken2018}. The analyst's task is comparing the probability that $x$ and $y$ have a common source given $x$, $y$, and the information $I$, i.e., $P(E_0|x,y,I)$, to the probability that $x$ and $y$ have a different source, given the same information, $P(E_0^\mathsf{c}|x,y,I)$.

After being given the evidence and other case information, the analyst would like to update his belief concerning the event $E_0$ in a rational and coherent manner. Denoting the analyst by $A$, this task can be written by using the odds form of Bayes' rule:
\begin{align}
\label{eq:bayesruleanalyst}
\underbrace{\frac{P(E_0)}{P(E_0^\mathsf{c})}}_{\text{Prior odds}_{A}} \times \underbrace{\frac{P(x,y,I|E_0)}{P(x,y,I|E_0^\mathsf{c})}}_{\text{Likelihood ratio}_{A}}  = \underbrace{\frac{P(E_0|x,y,I)}{P(E_0^\mathsf{c}|x,y,I)}}_{\text{Posterior odds}_{A}}.
\end{align}
The prior odds is the analyst's beliefs about $E_0$ prior to seeing the evidence, the likelihood ratio is the quantity at which the analyst arrives through the analysis of the evidence, and the posterior odds is the final quantity at which the analyst will arrive after rationally incorporating all available information. 

Ideally, the analyst should hold prior odds that are in some sense ``neutral,'' so that their decision will be made entirely based on the evidence contained in the data. In our setting, we assume that the analyst believes there is a set of $N$ suspects, and starts with the belief that each has an equal chance of being the source of the latent print. In that case, the prior odds are given by $1/N$. 

The analyst updates their prior with new information via the likelihood ratio. Likelihood ratios that are greater than one indicate that the observations $x$ and $y$ are more likely when $S_0$ is the source of both $x$ and $y$ (i.e. the event $E_0$), and small likelihood ratios (between zero and one) arise when $x$ and $y$ are more likely when $S_0$ is not the source of the crime scene evidence (i.e. the event $E_0^c$). We follow \cite{lundiyer} in assuming that the denominators of these three are never zero, which implies that the analyst never takes a ``hardline'' or extreme stance.

The likelihood ratio has a long history in forensic science, dating to at least \citet{lindley}, who used it to evaluate whether two glass fragments came from the same source. It is now common practice for forensic analysts to report a likelihood ratio for single-source DNA \citep{national1996evaluation, stockmarr_likelihood_1999, steele2014dna}, rather than simply stating that the questioned sample came from a specific source. Although there has been a push for fingerprint analysts to state their conclusions as likelihood ratios in the United States \citep{abraham2013modern}, source conclusions for fingerprint evidence are most often reported as categorical conclusions: ``identification'' (same-source), ``exclusion'' (different sources), or ``inconclusive''. Reporting the value of evidence using the likelihood ratio has gained traction in Europe and internationally \citep{ensfireport}. See \citet{Aitken2018} for a more detailed overview of the use of the likelihood ratio to evaluate evidence. Through the remainder of this article, we will use the likelihood ratio to quantify the value of evidence, but recognize that it is not the only way that the value of evidence is expressed \citep{kayeblog}. We also note that our results are not dependent on how the likelihood ratio is calculated (e.g., using a Bayesian or classical approach to statistics) and generalizes to different settings.

In order to carry out a fair analysis for all suspects, forensic analysts should start with the belief that each of the $N$ suspects has the same probability of having produced the crime scene print. In other words, the analyst should use the principle of indifference to determine the probability that each of the $N$ suspects left the print at the crime scene. Thus forensic analysts should have a prior odds of 
\begin{equation}
\label{eq:priorodds1}
\frac{P(E_0|I)}{P(E_0^\mathsf{c}|I)} = \frac{1/N}{N-1/N} = \frac{1}{N-1},
\end{equation}
or equivalently (if we just say there are N+1 suspects without changing the meaning of the equation),
\begin{equation}
\frac{P(E_0|I)}{P(E_0^\mathsf{c}|I)} = \frac{1/(N+1)}{N/(N+1)} = \frac{1}{N}.
\end{equation}

However, it is possible that the analyst has incorrect prior beliefs either because his prior odds are not $1/N$, or because the analyst is confusing the version of the prior odds before conditioning on $I$,---$P(E_0)/P(E_0^\mathsf{c})$---with the version in which the terms are conditioned by $I$,---$P(E_0|I)/P(E_0^\mathsf{c}|I)$. It is possible that
\begin{equation}
\label{eq:priorodds2}
\frac{P(E_0)}{P(E_0^\mathsf{c})} = 1/N, \quad \text{but} \quad \frac{P(E_0|I)}{P(E_0^\mathsf{c}|I)}\neq 1/N.
\end{equation}
Thus, as soon as the analyst learns $I$, his prior odds change from the fair $N$ to something else.

We know that the analyst can be biased by information about his base rate (i.e. his observed distribution of identifications, exclusions, and inconclusives), education and training, personal factors, and cognitive factors \citep{drorpyramid}. As the analyst gains experience in casework, if he does not receive ground-truth feedback about whether he arrived the correct conclusion, he will acquire long-run bias in his prior distribution \citep{drorpyramid}. We might instead assume the analyst uses their previous experience as a prior, taking the share of prints determined to have come from the same source that they've been shown as their prior odds. So if, for example, out of all of the prints they've seen so far in their careers they have concluded the prints came from the same source 20\% of the time, their prior will be 0.2. Finally, the analyst might have some beliefs in the ability of the police (or other customer) to deliver the correct suspect to them, and may take their prior odds to be their estimation of the abilities of the police. They might imagine, for example, a set of $N$ possible perpetrators, but think the suspect identified by the police is more likely than the others to be the true source of the crime scene prints. In our analysis, we will assume the prior odds of $1/N$, as the most neutral of these options. 

We take a different approach to describing the quantity of interest than other researchers \citep{lundiyer} since we propose that the analyst should arrive at a posterior odds, not just a likelihood ratio. While we agree that the analyst should calculate a likelihood ratio, we also think that since they are rational actors, they arrive at specific posterior conclusions after learning this likelihood ratio, represented by the posterior odds. These odds can then be presented to the trier of fact for them to make a decision about guilt. 

If the posterior odds is greater than one, the evidence suggests that $x$ and $y$ have the same source; if it is between zero and one, the evidence suggests that someone else is the source of $y$. If the posterior odds is near one, neither conclusion is favored by the evidence. What should and should not be included in the information $I$?

\subsection{Task-relevant and task-irrelevant information for the analyst's task}\label{sec:taskrelevanttaskirrelevant}

Should the analyst be told every piece of information that is known to the police, and to other analysts, about the case? Task-relevant and task-irrelevant information are two types of information that distinguish between what should be known by the analyst and what should not \citep{nationalcommission}. It is useful to define these two types of information by using conditional independence statements \citep{kayeblog}.

Information is task-relevant if and only if $(x,y) \not\!\perp\!\!\!\perp I \mid E_0$, which is equivalent to 
\begin{equation}
P(x,y \mid E_0) \neq P(x,y \mid I, E_0).
\end{equation}
Information is task-irrelevant if and only if $(x,y)\ci I \mid E_0$, which is equivalent to
\begin{equation}
P(x,y \mid E_0) = P(x,y \mid I, E_0).
\end{equation}
In each case $E_0$ can be replaced with $E_0^\mathsf{c}$ and the definitions still hold.

Intuitively, information is task-relevant if it has the potential to assist the analyst in evaluating the probability of observing the two pieces of evidence if they really are from the same source. It is task-irrelevant if it does not help the analyst in their task if they are from the same source. 

\begin{figure}[ht]
   \centering
   \includegraphics[width=\textwidth]{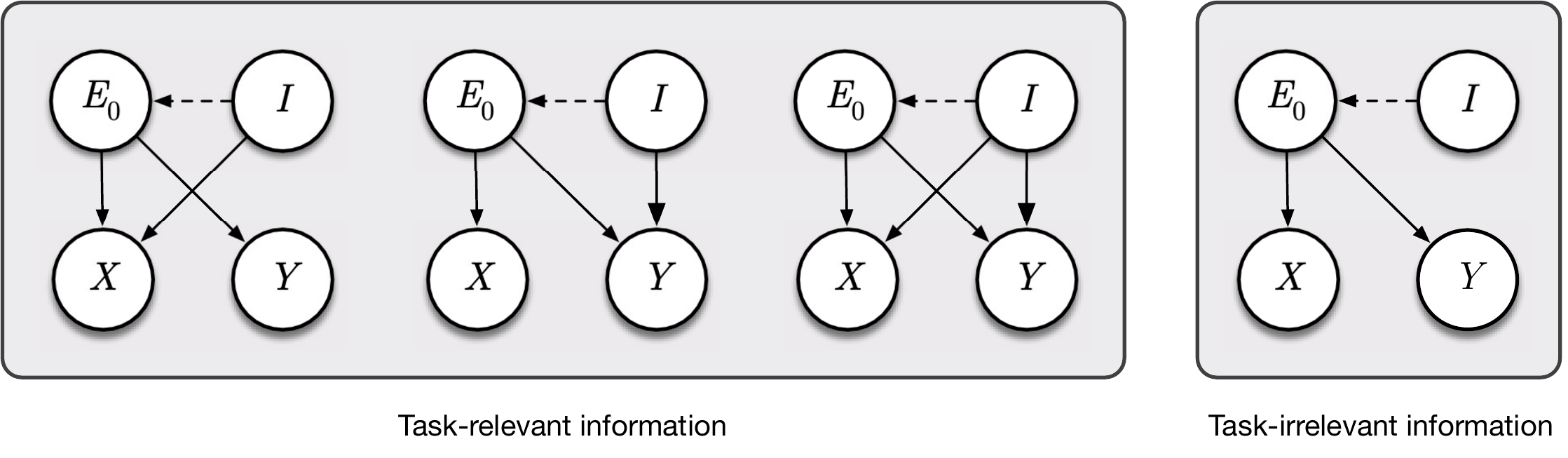} 
   \caption[Directed acyclic graphs showing examples of task-relevant and task-irrelevant information.]{Directed acyclic graphs (DAGs) showing of task-relevant and task-irrelevant information. The edges represent causal influence. The dashed line indicates that the edge can be present or absent and the definition still holds.}
   \label{fig:dag}
\end{figure}

A more precise way to understand the difference between task-relevant and task-irrelevant information is by using a directed acyclic graph. Fig. \ref{fig:dag} shows a graphical representation of examples of task-relevant and task-irrelevant information. Intuitively, if $I$ is task-irrelevant, then if the pair of prints ($x,y$) was truly left by the same individual, knowing $I$ does not change the likelihood of observing the evidence. If it is task-relevant, knowing $I$ does affect it.

Task-relevant information often affects the physical appearance of the evidence. For example, if an individual's fingerprint was found on a curved surface, it will appear wider than a print left by the same individual on a flat surface. In Fig. \ref{fig:dag}, this can be shown by using the left-most DAG. $I$ is the fact that the print $X=x$ was left on a curved surface while $Y=y$ was not. $I$ thus affects both how $x$ looks and whether the prints were left by the same individual.

On the other hand, task-irrelevant information is often unrelated to the physical appearance of the evidence. For example, a suspect's criminal history is most likely task-irrelevant to a forensic comparison task. In Fig. \ref{fig:dag}, this can be shown by using the right-most DAG. The reason is that the likelihood two fingerprints ($x,y$) are similar does not depend on the suspect's criminal history ($I$) except through the fact that someone with a criminal history is more likely to leave a print at the scene ($E_0$). So, it's irrelevant to the task of the analyst. 

\subsection{The trier of fact's task}

It is not the role of the forensic analyst to determine the ultimate guilt of the suspect, but the role of the ``trier of fact'' (TF) -- a decision-maker who makes a final determination of guilt, such as a judge or jury. In legal trials, the trier of fact is a person (e.g. judge), or group of people (e.g. jury), who determines factual issues in a legal proceeding, meaning that they decide, from the evidence, whether something existed or some event occurred. In reality, the types of decisions vary depending on the case and jurisdiction. For simplicity in our exposition, we define the trier of fact as the entity responsible for deciding whether an individual is guilty or not guilty of a crime. 

We can decompose the posterior odds held by the trier of fact. We define the evidence seen by the trier of fact as $\boldsymbol{M} = \{M_1, \cdots, M_m, I\}$ where each $M_j$ is the evidence reported by an analyst, police, or others in the courtroom. So for example, if $\boldsymbol{M}$ is all the evidence in a murder case, $M_1$ may be a statement of ``identification'' from the fingerprint analyst, $M_2$ may be a statement from the firearms analyst, and so on. Note that all the $M_j$'s are functions of the posterior odds ratio estimated by each analyst. In addition, the trier of fact must decide whether the defendant(s) are either guilty ($G$) or not guilty ($G^{\mathsf{c}}$). The trier of fact version of Bayes' rule in ratio form is therefore
\begin{equation}
\underbrace{\frac{P(G)}{P(G^{\mathsf{c}})}}_{\text{Prior odds}_{\text{TF}}} \times 
\underbrace{\frac{P(\boldsymbol{M} \mid G)}{P(\boldsymbol{M} \mid G^{\mathsf{c}})}}_{\text{Likelihood ratio}_{\text{TF}}}
= \underbrace{\frac{P(G\mid \boldsymbol{M})}{P(G^{\mathsf{c}}\mid \boldsymbol{M})}}_{\text{Posterior odds}_{\text{TF}}}.
\end{equation}

Thus, if the posterior odds of the trier of fact are high, then the verdict will be guilty. The exact value that this ratio requires to represent a guilty verdict is left up to the trier of fact to decide. It is possible for a trier of fact to introduce new biases, e.g. jury selection. However, for this article, we assume the trier of fact does not introduce new biases and we focus on the biases introduced by the forensic analyses.

\section{Biases at the analyst level}\label{sec:biasesattheanalystlevel}

In this section, we study how biases arise at the analyst level. We present a formalization of contextual bias, two examples (a hypothetical one and an empirical one), and then a formalization of what psychologists have called reference bias (we call it imputation bias) and how long-term effects of bias affect an analyst's prior. 

\subsection{Contextual bias formalization}

If we assume a neutral prior, bias will enter into the determination by the analyst through the likelihood. Note that this likelihood can be broken apart into two parts,
\begin{align}
    \underbrace{\frac{P(x,y,I|E_0)}{P(x,y,I|E_0^\mathsf{c})} }_{\text{Likelihood ratio}}
    &= 
    \underbrace{ \frac{P(x,y|I,E_0)}{P(x,y|I,E_0^\mathsf{c})}}_{\text{Part 1}} 
    \times 
    \underbrace{
    \frac{P(I\mid E_0)}{P(I\mid E_0^\mathsf{c})}}_{\text{Part 2}}
    \label{eq:likelihoodBreakdown}
\end{align}

Part 1 denotes the part of the likelihood that includes a comparison of how likely it is to observe the pieces of evidence given the additional information and the fact that they are same-source or different-source. Part 2 captures how the analyst acquires contextual bias. It denotes the belief the analyst has about the chances of seeing information $I$ depending on whether or not $x$ and $y$ have a common source. We assume in this case that at least part of $I$ is task-irrelevant.

Let us denote by $\alpha$ the share of suspects with characteristic $I=1$ among all those suspects who truly left a print at the scene over a long series of investigations. $\alpha$ is thus a true long-run approximation of $P(I=1 \mid E_0)$. Let us denote by $\beta$ the share of suspects with characteristic $I$ among those included in the suspect set who did not leave the print, a true approximation of $P(I=1 \mid E_0^\mathsf{c})$. We then have that part 2 is,
\begin{equation}\label{eq:biasalphabeta}
    \frac{P(I=1\mid E_0)}{P(I=1\mid E_0^\mathsf{c})} =  \frac{\alpha}{\beta}. 
\end{equation}
In other words, $\alpha$ is the prevalence of trait $I$ within criminals and $\beta$ is the prevalence of that trait within non-criminals.

The analyst now updates his estimate of $P(I\mid E_0)$ according to past experience, external attitudes, or other evidence. Letting $\alpha^*$ be the analyst's beliefs about the chances that information $I$ is true given that the prints match, and $\beta^*$ be the analyst's beliefs about the chances that information $I$ is true given that the prints do not match. Then, $P_A$, the analyst's subjective probability distribution, is similar to \eqref{eq:biasalphabeta} above, but replacing true values with beliefs,
\begin{align}
    \frac{P_A(I=1 \mid E_0)}{P_A(I=1\mid E_0^\mathsf{c})} &=  \frac{\alpha^*}{\beta^*}.
\end{align}
Clearly, as long as $\frac{\alpha^*}{\beta^*} \neq \frac{\alpha}{\beta}$, this part of the likelihood will be biased since,
\begin{equation}
    \frac{P_A(I \mid E_0)}{P_A(I \mid E_0^\mathsf{c})} =
    \left\{
        \begin{array}{rc}
                \Big( \frac{\alpha^*}{\beta^*} \frac{\beta}{\alpha} \Big)
                \frac{P(I=1 \mid E_0)}{P(I=1 \mid E_0^\mathsf{c})} & \mathrm{if\ } I=1 \\
                \Big( \frac{1-\alpha^*}{1-\beta^*} \frac{1-\beta}{1-\alpha} \Big)
                \frac{P(I=0 \mid E_0)}{P(I=0 \mid E_0^\mathsf{c})}  & \mathrm{if\ } I=0. \\
        \end{array} 
    \right.
\end{equation}
Denote the bias term with $\delta_{I}(P_A,I)$ and we can rewrite this more simply as
\begin{align}
    \frac{P_A(I \mid E_0)}{P_A(I \mid E_0^\mathsf{c})} &= \delta_{I}(P_A,I) \frac{P(I \mid E_0)}{P(I \mid E_0^\mathsf{c})}.
\end{align}
Plugging this back into \eqref{eq:likelihoodBreakdown}, we see that if the analyst is given task-irrelevant information $I$ along with the evidence samples $x$ and $y$, the analyst's likelihood ratio will be biased by a multiplicative factor of $\delta_I$,
\begin{align}
    \underbrace{\frac{P_A(x,y,I|E_0)}{P_A(x,y,I|E_0^\mathsf{c})} }_{\text{Likelihood ratio}_{A}}
    &= 
     \delta_I(P_A, I) 
    \underbrace{\frac{P(x,y,I|E_0)}{P(x,y,I|E_0^\mathsf{c})} }_{\text{True likelihood ratio}}.
    \label{eq:contextual-bias}
\end{align}

\subsubsection{Hypothetical application: Race as task-irrelevant contextual information for a fingerprint comparison}\label{sec:raceexample}

How could knowing the suspect's race affect an analyst's performance in comparing two fingerprints? It is likely that most forensic analysts are not intentionally racist. Nevertheless, learning the suspect's race, a piece of task-irrelevant information, could lead an analyst to incorrect results. 

Suppose the supervisor tells the analyst that the suspect's race is Black. Then, $\alpha$ is the prevalence of Black individuals within the pool of criminals and $\beta$ is the prevalence of Black individuals within non-criminals. The bias $\delta_{I}(P_A,I)$ is a function of the beliefs held by the analyst encoded in $P_A$ and the contextual information about race, $I$. We might imagine that $\alpha^*$ is roughly estimated as the share of all suspects eventually found guilty who are Black. Holding $\beta=\beta^*$, as long as $\alpha^*>\alpha$, information about the race of any Black suspect will increase the estimated posterior odds.

Imagine that in reality $\alpha=\beta=P(I=1)$, meaning the chances of being Black given the suspect committed the crime are the same as the chances of being Black given they did not: both are simply the chances of being Black among the suspect set. The analyst, however, believes instead that Black individuals are overly represented among same-source suspects by a factor of two: $\alpha^* = 2P(I=1)$ (assume for convenience $P(I=1)<1/2$), and $\beta^* = \beta$. He might have this misconception from not having observed enough realizations of that type of evidence, lack of ground truth corrections, from media exaggerations, cultural biases, or other reasons.

In the neutral scenario we would expect to find that
\begin{equation}
\frac{P(I=1\mid E_0)}{P(I=1 \mid E_0^\mathsf{c})} = \frac{P(I=0\mid E_0)}{P(I=0 \mid E_0^\mathsf{c})}=1.
\end{equation}
We have that $\alpha^* = P_A(I=1\mid E_0) = 2P(I=1)$ and $\beta^* = P_A(I=1\mid E_0^C) = P(I=1)$. Thus,
\begin{equation}
    \frac{P_A(I=1 \mid E_0)}{P_A(I=1 \mid E_0^\mathsf{c})} = \frac{2P(I=1)}{P(I=1)} = 2,
\end{equation}
and
\begin{align}
    \frac{P_A(I=0 \mid E_0)}{P_A(I=0 \mid E_0^\mathsf{c})} & = \frac{1-\alpha^*}{1-\beta^*} \nonumber \\ 
    &= \frac{1-2P(I=1)}{1-P(I=1)} \nonumber \\ 
    &= \frac{2(1-P(I=1))-1}{1-P(I=1)} \nonumber \\
    &= 2 - \frac{1}{1-P(I=1)}.
\end{align}
The analyst's mistaken beliefs instead give
\begin{equation}\label{eq:alpha2}
    \frac{P_A(I=1 \mid E_0)}{P_A(I=1 \mid E_0^\mathsf{c})}  = 2 \cdot
    \frac{P_A(I=0 \mid E_0)}{P_A(I=0 \mid E_0^\mathsf{c})} = 2-\frac{1}{P(I=0)},
\end{equation}
which means that the bias factor due to learning the race $I$ is
\begin{equation}
    \delta_{I}(P_A,I) = 2 - \frac{1-I}{P(\text{Race} \neq \text{Black})}.
\end{equation}
Even if the crime scene ($x$) and suspect ($y$) data are correctly analyzed, the analyst may have posterior odds that are double their true value since the analyst's likelihood ratio is now twice what it should be
\begin{align}
    \frac{P_A(E_0\mid x,y,I=1)}{P_A(E_0^\mathsf{c}\mid x,y,I=1)} &= \frac{P(x,y|I=1,E_0)}{P(x,y|I=1,E_0^\mathsf{c})} \times \frac{2}{N}\\
    &= 2 \cdot \frac{P(E_0\mid x,y,I=1)}{P(E_0^\mathsf{c}\mid x,y,I=1)}.
\end{align}

This misspecification could severely increase the estimated posterior odds that the suspect committed the crime, even though this is incorrect. Note that this type of bias is entirely separate from the prior odds: even if the analyst were reporting the likelihood ratio and not the posterior odds, the likelihood ratio has doubled.

Given the physical evidence, it may be that suspects with characteristic $I$ are more likely to have produced highly similar prints because they are more likely to commit crimes. This would seem to argue for including $I$ in the interpretation of evidence, but we have shown that doing so can have negative consequences. Note that as more pieces of task-irrelevant information are given to the analyst in $I$, the bias will continue increasing. That is, there is no correction mechanism that will prevent the bias from growing after a certain amount of task-irrelevant information is incorporated into the analyst's judgment. 

We selected the example including the suspect's race here to show how misuse of this information, even unintentionally, could lead to biased judgments. Other protected characteristics, such as age, disability, sex, or religion could lead to biased judgments in much the same way as race. Examiners will probably agree that the suspect's protected characteristics are irrelevant the task of comparing fingerprints, yet learning this information can have a negative effect. Furthermore, our argument from this section holds for other non-protected characteristics that are task-irrelevant, such as whether the defendant has a prior criminal history. Note that in this example we are not saying that white and Black individuals are equally likely to commit a crime. We are saying that the two prevalences $\alpha$ and $\beta$ are equal to the population prevalence. We still find bias, even if $\alpha$ is not equal to $\beta$, i.e., even if there is a true difference in the prevalence between criminals and non-criminals.

\subsubsection{Real-world application: The case of Brandon Mayfield and the Madrid Bombing}

It is difficult to find evidence of bias in forensic analyses because (a) casework conclusions are rarely made public outside of a courtroom, and (b) simulating real-world circumstances for an experiment is challenging. A notable exception is the infamous FBI error in the Mayfield case described in the Introduction, which was publicly reported on and allowed researchers to perform an experiment to see if contextual bias could be detected \citep{dror-mayfield}.  

The researchers selected five experienced examiners who 1) knew about the Mayfield case but 2) had not examined the prints from that case. For each examiner, the researchers 3) selected a pair of prints that the examiner had concluded was a match (identification) some time earlier. Then, one of their colleagues asked them to review the prints, but 4) told the examiner that the prints were from the Mayfield case (``thus creating an extraneous context that the prints were a non-match''). Three of the examiners changed their mind from match to non-match (exclusion), one changed to ``can't decide,'' and one did not change. This, the authors argue, is evidence that ``Contextual information renders experts vulnerable to making erroneous identifications.''

We can formalize what happened in this experiment within the framework described in Equation \ref{eq:contextual-bias}. For an analyst's bias $\delta_I$ due to their beliefs $P_A$, and the contextual information $I$ from 4) above, the analyst has biased likelihood ratio,
\begin{align}
    \frac{P_A(E_0\mid x,y,I)}{P_A(E_0^\mathsf{c}\mid x,y,I)}
    &= \delta_I(P_A, I) \cdot \frac{P(E_0\mid x,y,I)}{P(E_0^\mathsf{c}\mid x,y,I)},
\end{align}
and biased posterior odds as well,
\begin{align}
    \frac{P_A(x,y,I=1 \mid E_0)}{P_A(x,y,I=1 \mid E_0^\mathsf{c})} &= \delta_I(P_A, I) \frac{P(x,y,I=1 \mid E_0)}{P(x,y,I=1 \mid E_0^\mathsf{c})}.
\end{align}
We do not know the values of $\delta_I$ for each examiner exactly, but given the results from \citet{dror-mayfield}, one plausible situation is the following: For the three examiners who switched from ``match'' to ``no match'', $\delta_I=2$, for the one who switched to ``can't decide'', $\delta_I=1.5$, and for the one who did not switch, $\delta_I=1$. Thus, the average bias for the group can be calculated as the average of the individuals' biases, 
\begin{equation}
    \bar{\delta_I} = \frac{3 + 3 + 3 + 1.5 + 1}{5} = 1.7.
\end{equation}
This is the average bias due to the examiners being told that the prints were from the Mayfield case (a task-irrelevant piece of information). Of course, the experiment had a small sample size of five, and we could obtain more precise estimates of $\bar{\delta_I}$ from a larger sample. 

\subsection{Imputation bias}

Imputation bias refers to the situation in which the analyst uses information from the reference print to mark up the crime scene print. This is done unintentionally and likely unconsciously, without malice. The task-irrelevant information in this case is contained within the physical evidence given to the analyst, not external, contextual information. This has sometimes been called data-bias and reference-material-bias \citep{drorpyramid}.

\begin{figure}[ht!]
   \centering
   \includegraphics[width=.7\textwidth]{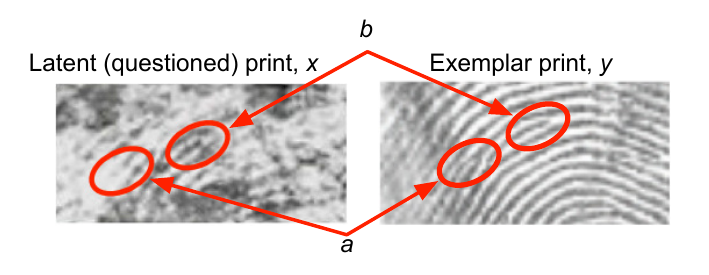} 
   \caption{Latent print from the Madrid bombing crime scene (left), and exemplar print from suspect (right). It is easy to see that the minutiae at $b$ correspond, but it is unclear whether the minutiae at $a$ correspond as well since $x$ is missing information.}
   \label{fig:imputationbiasmayfield}
\end{figure}

The suspect's exemplar print is often a clean print taken under controlled circumstances, but the latent print has low quality, meaning it is missing information in some areas. See, for example, the prints in Fig. \ref{fig:imputationbiasmayfield}. The latent print from the Madrid bombing crime scene (left) is missing information in some areas, while the exemplar print from Brandon Mayfield (right) is more complete. The locations marked by $b$ clearly point to corresponding minutiae (ridge endings) in both prints, but for locations $a$ it is not clear whether there are corresponding minutiae because the latent print is missing information.

\begin{figure}[ht]
    \centering
    \includegraphics[width = \textwidth]{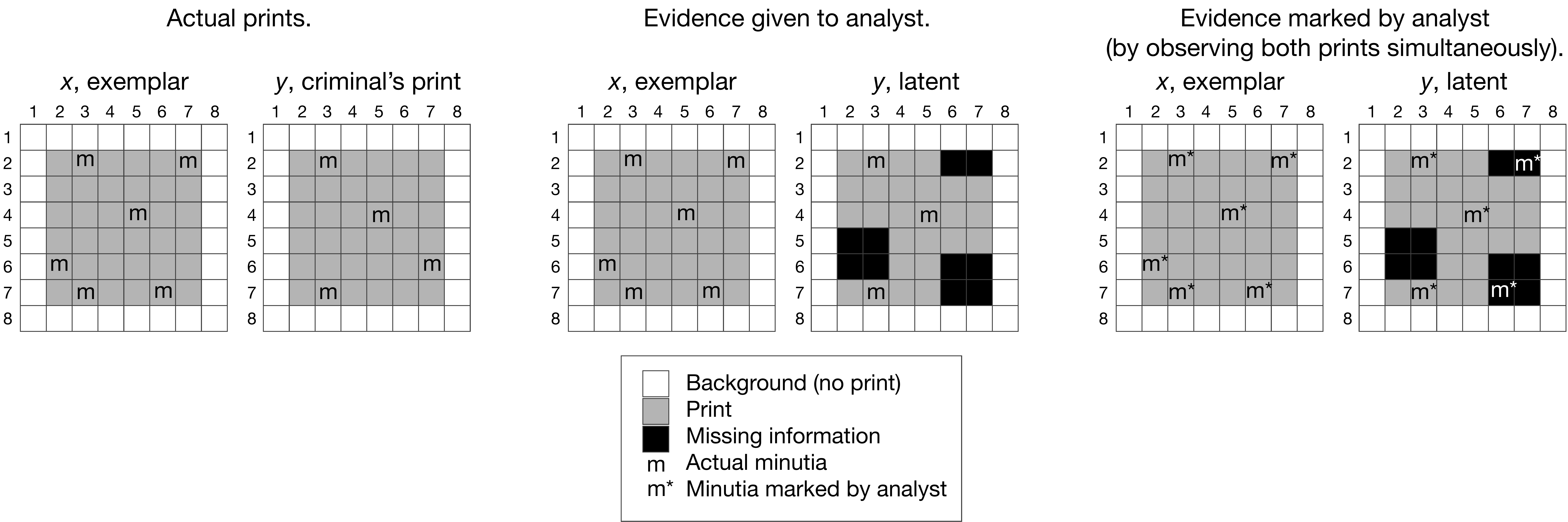}
    \caption{Example of imputation bias. The actual prints $x$ and $y$ have minutiae in certain places (left pair). The analyst only observes $y$ with missing information (central pair), and assumes that the minutiae in $x$ is also in $y$, incorrectly (right pair).}
    \label{fig:imputationbias}
\end{figure}

Fig. \ref{fig:imputationbias} illustrates how imputation bias could occur. The left-most pair of prints are the actual prints, which the analyst does not observe and has actual minutiae (denoted with m's). The middle pair is the evidence given to the analyst, and the right-most pair is the analyst's marked-up prints, with minutiae marked by the examiner (m$^*$). In the middle pair, the areas with missing information (black squares) could be marked as containing minutiae by the analyst. It makes sense that if a gray square right next to a black square looked like a minutia was beginning, it would end inside the black square. Since the exemplar and the latent have so many minutiae in common already, the analyst might think that the minutiae in $x$ should also be in $y$. However, it could be that \textit{precisely} those were not in $y$, as can be seen in the actual prints. Thus, the minutiae in the marked-up print differs from the actual print, and this may lead to a false-positive identification error.

We now demonstrate how imputation bias can be generated within the likelihood ratio. We break down the likelihood even further from Equation \eqref{eq:likelihoodBreakdown}, so that,
\begin{align}
    \underbrace{\frac{P(x,y,I|E_0)}{P(x,y,I|E_0^\mathsf{c})} }_{\text{Likelihood ratio}_{A}}
    &= 
    \underbrace{ \frac{P(y|x,I,E_0)}{P(y|x,I,E_0^\mathsf{c})}}_{\text{Part I.1}} 
    \times
    \underbrace{ \frac{P(x|I,E_0)}{P(x|I,E_0^\mathsf{c})}}_{\text{Part I.2}} 
    \times 
    \underbrace{
    \frac{P(I\mid E_0)}{P(I\mid E_0^\mathsf{c})}}_{\text{Part II}}
    \label{}
\end{align}
We already saw how problems with Part II could arise in section \ref{sec:biasesattheanalystlevel}. Here we are interested with Part I.1. This gives the probability distribution of minutiae on print $y$ given all the other information, $x,I,E_0$ and $x,I,E_0^\mathsf{c}$. 

To simplify the example, we can encode $y$ as a vector of 0's and 1's, where $y_i=1$ means there is minutiae in position $i$ of the print. The same holds for the vector $x$. The task can then be conceptualized as determining how likely it is to see two vectors with the observed degree of similarity. For example, we might have the following data where out of six positions on the print, we see four matches, two of which are matching minutiae. The question is now what are the chances of four matches given the two samples ($x$ and $y$) come from the same source, and what are the chances of four matches given they come from different sources. This is a very abstracted version of the task analysts actually perform, and is meant only for illustration of the probability problem. 

\begin{table}[ht!]
    \centering
    \begin{tabular}{c|c|c}
         $x$ & $y$ & Correspond\\
         \hline
         0 & 0 & 1\\
         0 & 1 & 0\\
         0 & 0 & 1\\
         1 & 1 & 1\\
         1 & 1 & 1\\
         1 & 0 & 0\\
    \end{tabular}
    \caption{Example of simplified minutiae tagging task. The zeroes and ones correspond to the presence or absence of a minutia, and this supposes that there is only one type of minutia.}
    \label{tab:MinutiaeTable}
\end{table}

Having tagged minutiae in both prints, the analyst might have some function $f(x,y)$ to come to their source conclusion. For example, in this extremely simplified case, they might say,
\begin{equation}
    f(x,y) =
    \left\{
        \begin{array}{cc}
                \text{Source Identification} & \mathrm{if\ } \sum_{i=1}^m \indc{y_i = x_i = 1} \geq 12 \\
                \text{Support for Same Source} & \mathrm{if\ } 7 \le \sum_{i=1}^m \indc{y_i = x_i = 1} < 12 \\
                \text{Inconclusive} & \mathrm{if\ } 3 \le \sum_{i=1}^m \indc{y_i = x_i = 1} < 7 \\
                \text{Exclusion} & \mathrm{otherwise}.
        \end{array} 
    \right.
\end{equation}

If some of the information in a section is missing or blurry, the analyst's task to tag minutiae accurately becomes much more difficult. Imagine if instead of the information in Tab. \ref{tab:MinutiaeTable}, the analyst could only observe the information in Tab. \ref{tab:MinutiaeTableMissing}, where a missing value is represented by a question mark.

\begin{table}[ht!]
    \centering
    \begin{tabular}{c|c|c}
         $x$ & $y$ & Correspond\\
         \hline
         0 & ? & ?\\
         0 & ? & ?\\
         0 & 0 & 1\\
         1 & ? & ?\\
         1 & 1 & 1\\
         1 & 0 & 0\\
    \end{tabular}
    \caption{Example of simplified minutiae tagging task with missing information.}
    \label{tab:MinutiaeTableMissing}
\end{table}

The analyst must now make decisions about where to tag minutiae with this limited dataset. One danger is that the analyst--likely inadvertently--uses information from $x$ or $I$ to ``impute'' the missing values of $y$. 

Imputation is a strategy to fill in missing data by learning a model from observed data and using that to predict missing values of a variable. Imagine the analyst fills in each missing bit of information with information from $x$, giving,
\begin{equation}
    y_{i}^* =
    \left\{
        \begin{array}{cc}
                y_{i} &\text{if } y_{i} \text{ observed} \\
                x_{i} &\text{if } y_{i} \text{ not observed} 
        \end{array} 
    \right.
\end{equation}

This results in the dataset shown in Tab. \ref{tab:MinutiaeTableImputed}.
\begin{table}[ht!]
    \centering
    \begin{tabular}{c|c|c}
         $x$ & $y*$ & Correspond\\
         \hline
         0 & 0 & 1\\
         0 & 0 & 1\\
         0 & 0 & 1\\
         1 & 1 & 1\\
         1 & 1 & 1\\
         1 & 0 & 0\\
    \end{tabular}
    \caption{Example of simplified minutiae tagging task with missing pixels imputed.}
    \label{tab:MinutiaeTableImputed}
\end{table}

We see that in this case, the analyst has mistakenly declared the second row to be an area of correspondence between the latent and reference print.

This is obviously an extreme example, but we might easily imagine that an analyst observing two prints might come to believe they see minutiae in a blurry section of $y$ after having seen minutiae in the corresponding section of $x$, especially if they have other reasons to believe the prints are from the same source. After this imputation process, $y$ has been replaced with $y^*$, which necessarily looks more like $x$ than the original $y$. Thus,
\begin{align*}
    P(y^*\mid E_0,I,x) &\geq P(y\mid E_0,I,x) \text{  and}\\
    P(y^*\mid E_0^\mathsf{c},I,x) &\leq P(y\mid E_0^\mathsf{c},I,x)
\end{align*}
Because of this, the new likelihood will be inflated relative to the truth, by some bias amount we call $\delta_{\text{Impute}}(P_A,I,x,y)$,
\begin{align}
    \label{eq:LikelihoodBias}
    \underbrace{\frac{P_A(y\mid E_0,I,x)}{P_A(y\mid E_0^\mathsf{c},I,x)}}_{\text{Part I.1}_A} &= \frac{P(y^*\mid E_0,I,x)}{P(y^*\mid E_0^\mathsf{c},I,x)}\\
    &= \delta_{\text{Impute}}(P_A,I,x,y) \underbrace{\frac{P(y\mid E_0,I,x)}{P(y\mid E_0^\mathsf{c},I,x)}}_{\text{Unbiased Part I.1}_A}
\end{align}

This extreme model is thankfully unlikely, but imputation need not be absolute in order to cause similar issues. The issue arises as soon as the latent print is labeled using outside information, without altering the likelihood ratio calculation to account for this. Note that this section again does not depend on the inclusion of the prior.

\begin{figure}[ht]
    \includegraphics[width=\textwidth]{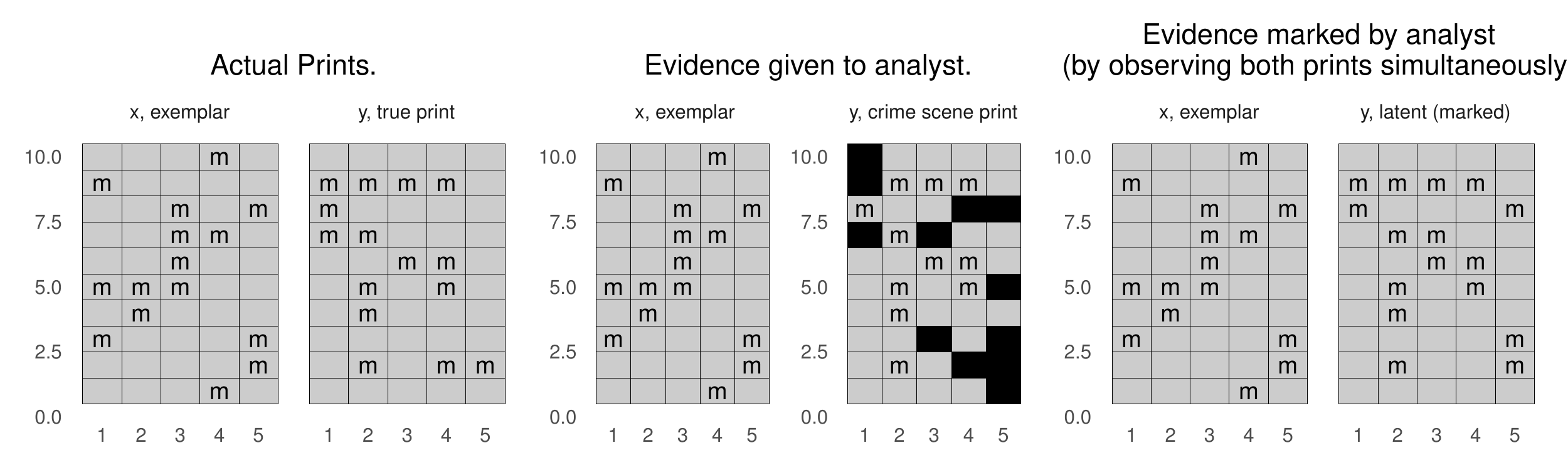} %
    \caption{Example of imputation bias, similar to Fig. \ref{fig:imputationbias}, but generated through simulation.}
    \label{fig:toyexample-imputation}%
\end{figure}

Now we illustrate imputation bias through a simple toy simulation. We can imagine fingerprints as 10x5 grids of information. Let $x$ and $y$ both be examples of such a grid, each with an expected 15 minutiae distributed uniformly across their surface. Marked presence of a minutia is represented as an $m$ and missing information is represented in black. The presence of minutiae for $x$ (Fig. \ref{fig:toyexample-imputation}(a), right) is perfectly known, while the true $y$ (Fig. \ref{fig:toyexample-imputation}(a), left) is unobserved. 25\% of the information in $y$ print is missing from the crime scene latent print (Fig. \ref{fig:toyexample-imputation}(b), right).

If the analyst could observe all the missing pixels (i.e. areas represented by boxes in the figures) in $y$, he would see Fig. \ref{fig:toyexample-imputation}(a), right. With imputation, the analyst replaces missing pixels of the latent $y$ with the corresponding pixels of $x$ and marks under these conditions \ref{fig:toyexample-imputation}(c), right. In this setting, the analyst mistakenly marks missing information as minutiae.

The true prints share 5 minutiae. The print with missing sections shares only 3 with the exemplar print, so if the analyst did no imputation, they would find only 3 matching minutiae. If, however, the analyst filled in the missing pixels with pixels from $x$, they would mark 8 matching minutiae. In this case, the conclusion of the analyst may switch from `exclusion' or `inconclusive' to `support for same-source'. It is clear that with the additional information, the suspect and crime scene prints appear much more similar, although in truth they are generated by completely different underlying patterns. 

\subsection{Long-term effects of bias in an analyst's prior}\label{sec:longterm}

We now return to the prior discussion of the forensic analyst's prior odds. In most cases, we can make the assumption that a biased prior will be corrected over time and with sufficient data by the likelihood, giving unbiased posterior odds. Unfortunately, this mechanism may not operate properly in some forensic disciplines. Suppose an analyst performs an examination and comes to a conclusion that a suspect's print and a crime scene print had a common source. Then the analyst testifies in court, and the trier of fact determines the suspect is guilty. The analyst then believes they have made a correct decision, since they agreed with the trier of fact. However, it is possible that the suspect was wrongfully convicted. Since the arrival of DNA evidence, it has become clear that many convictions were wrongful \citep{innocenceproject1, kennedy}. Will the fingerprint analyst correct their views once the suspect is found to have been wrongfully convicted? And what if an individual who is wrongfully convicted is never found to be so? It is estimated that about 6\% of individuals were wrongfully convicted. \citep{innocenceestimate}. It is likely that since the analyst often does not hear feedback about their conclusions, and especially if they take convictions as positive feedback, they will not correct their conclusions in the long run as they analyze more data over time. 

Assume that when an analyst sees a conviction, they believe that the defendant in fact did commit the crime and that there was no mistake in the conviction. Analysts will update their prior based on the conviction. Recall the context from earlier (Section \ref{sec:raceexample}), where the fingerprint analyst wrongly believes $P(I \mid E_0) = 2P(I)$. The analyst may also impute data in the likelihood and eventually decide that a suspect did leave the crime scene fingerprint. The initial bias of the fingerprint analyst may have been based on the fact that among all the suspects he had seen who were eventually found guilty, a large share were Black. This new conviction of a Black suspect increases this share even further. The more times we repeat this experiment, the more biased his prior odds become. 

Remember that we are concerned with situations where the analyst's prior beliefs about $\alpha$ are far from the truth. Let the true probability be given as
    \begin{align}
    P(E_0|I) &= \frac{\alpha}{P(I)(N+1)},
    \end{align}
and assume the analyst instead believes
    \begin{align}
    P_A(E_0|I) &= \frac{\alpha^*}{P(I)(N+1)}.
    \end{align}
For some $\alpha^* \neq \alpha$. Generally in these situations, the decision-maker will have a chance to observe many instances and update their beliefs. In forensic science settings, however, there are few repetitions of the decision-making scenario, and the ground truth may not be revealed to the analyst, or it may be revealed in a way which reinforces their mistaken beliefs. 

First, the ground truth of whether two prints were produced by the same source is rarely known. Analysts must update their prior beliefs using incomplete or even incorrect labels. Especially in forensic settings, one can imagine how this could introduce positive feedback loops, where suspects with certain characteristics are more often convicted, regardless of guilt. These past convictions are factored into the analyst's updates as true positives, reinforcing the problematic prior. 

This is shown via simulation below. We have a strong beta prior on $\alpha$ with a mean of 0.6. In the first figure, the analyst gets to see 100 correctly convicted suspects, and observe the share who have a prior criminal history. In the second, the analyst sees the same number of suspects, but some are incorrectly convicted, in a way that is biased. In this second case, the prior still moves towards the truth, but at a much slower rate than in the first case.

\begin{figure}[ht] \footnote{Plotting code from: https://alexanderetz.com/2015/07/25/understanding-bayes-updating-priors-via-the-likelihood/}
   \centering
   \includegraphics[width=.9\textwidth]{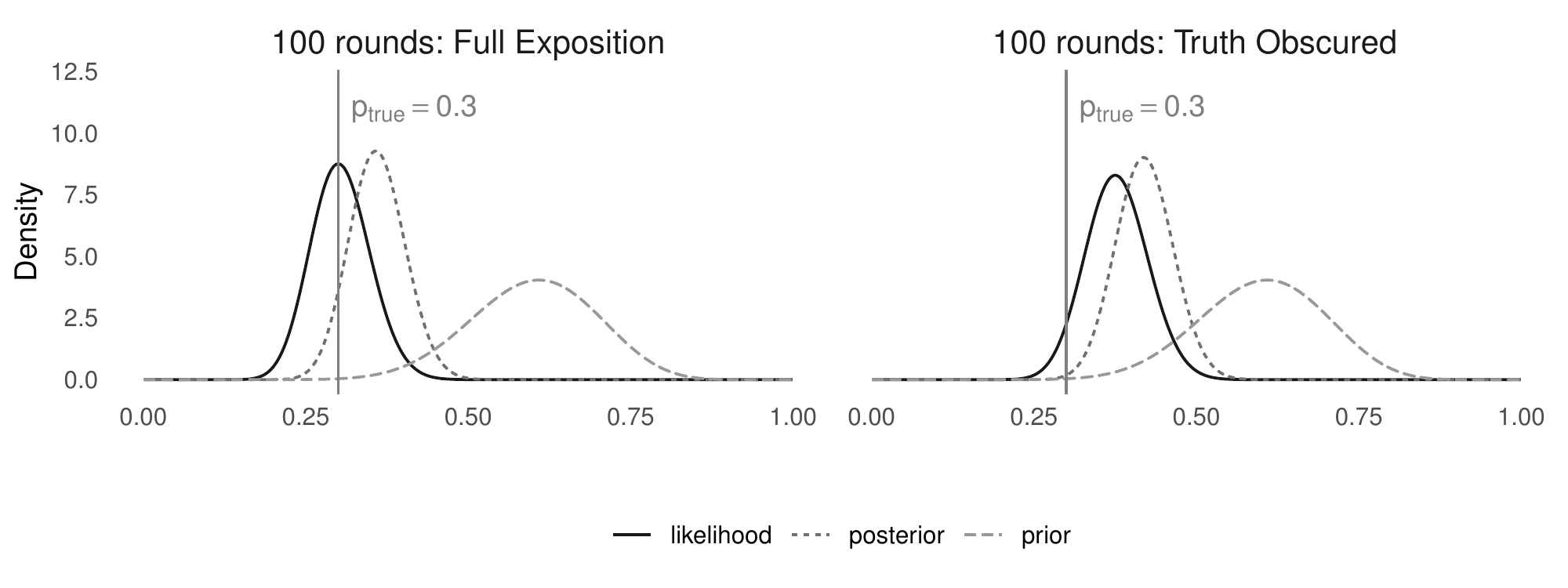} 
   \caption{The analyst's posterior estimate of $\alpha$ when he is allowed to see the truth (left) and when the truth is revealed in a biased way (right). The vertical line represents the true value.}
   \label{fig:obscuredPrior}
\end{figure}

Second, the analyst has access to a very limited number of examples. If their prior beliefs about the importance of prior criminal history (or race, sex, confessions, an alibi, etc) are sufficiently far from the truth, there will not be enough data to correct these beliefs. If we imagine that their prior beliefs about evidence are close to the true parameter, while those about prior criminal history are far from the true parameter, the posterior which withholds information about criminal history will be less prone to error. That is, a model which does not include criminal history and updates only on evidence will do better, after a relatively small number of steps, than a model which updates on both criminal history and evidence if the prior on criminal history is sufficiently wrong.

If this is the situation, information about suspects should be withheld from the analyst. Even an experienced technician will have trouble correctly estimating the necessary probabilities, which will bias the final results. Next, we discuss how bias spreads after it has affected an analyst's conclusions.

\section{How bias is propagated among analysts}

In this section, we focus on how bias is propagated among analysts throughout the laboratory. Organizational factors can lead to bias being propagated from one analyst to the next in a bias cascade, a bias snowball \citep{drorpyramid}, and bias in final determination of guilt. These organizational factors affect the likelihood ratio for the trier of fact.

\subsection{Bias Cascade}

\begin{figure}[ht]
   \centering
   \includegraphics[width=.8\textwidth]{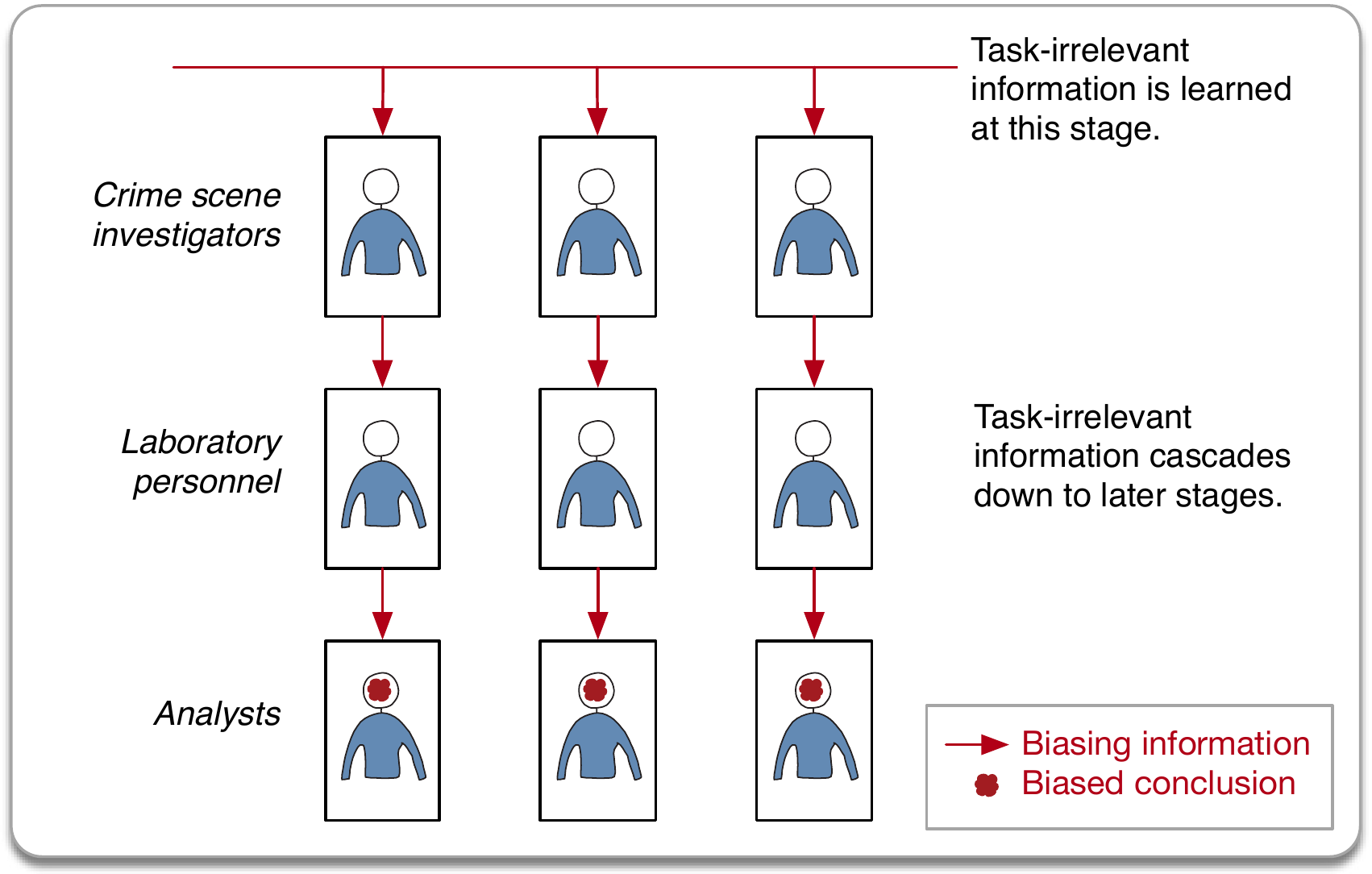} 
   \caption{Bias cascade example. At the crime scene, several investigators learn information that is task-irrelevant for analysts. This information is transferred to the laboratory, and then to the analysts. The analysts are all biased by this cascading of task-irrelevant information, even if they do not communicate with each other.}
   \label{fig:biascascade}
\end{figure}

The \textit{bias cascade effect} is when ``bias arises as a result of irrelevant information cascading from one stage to another, e.g. from the initial evidence collection to the evaluation and interpretation of the evidence'' \citep{dror2017letter}. Fig. \ref{fig:biascascade} shows that when analysts communicate with laboratory personnel, even if they are not given task-irrelevant information, this could lead to biased conclusions.

The analyst has allowed information about the suspect to enter into their decisions about the evidence -- either as they judged the chances of evidence coming from the same source given the suspect information or as they marked the evidence itself, or both. The final posterior is then pulled from the truth, despite the fact that the analyst has only learned true information. This bias is formalized as,

\begin{align*}
    \underbrace{\frac{P_A(E_0\mid I, x, y)}{P_A(E_0^\mathsf{c}\mid I, x, y)}}_{\text{Posterior Odds}_A} 
    &= \overbrace{\underbrace{\frac{P_A(y\mid x,I, E_0)}{P_A(y\mid x,I, E_0^\mathsf{c})}}_{\text{Part I.1}_A}
    \cdot
    \underbrace{\frac{P_A(x\mid I, E_0)}{P_A(x\mid I, E_0^\mathsf{c})}}_{\text{Part I.2}_A} 
    \cdot
    \underbrace{\frac{P_A(I\mid E_0)}{P_A(I \mid E_0^\mathsf{c})}}_{\text{Part II}_A}}^{LR_A}
    \cdot
    \underbrace{\frac{P_A(E_0)}{P_A(E_0^\mathsf{c})}}_{\text{Prior}_A}\\
    &= \delta_{\text{Impute}} \frac{P(y \mid x,E_0,I)}{P(y \mid x,E_0^\mathsf{c},I)} 
    \cdot
    \frac{P(x\mid I, E_0)}{P(x\mid I, E_0^\mathsf{c})} 
    \cdot 
    \delta_{I} \frac{P(I \mid E_0)}{P(I \mid E_0^\mathsf{c})} \frac{P(E_0)}{P(E_0^\mathsf{c})}\\
    &= \delta_{\text{Cascade}} \underbrace{\frac{P(E_0\mid I, x, y)}{P(E_0^\mathsf{c}\mid I, x, y)}}_{\text{Unbiased Posterior Odds}}.
\end{align*}

That is, our estimated posterior odds that the suspect and latent print come from the same source is $\delta_{\text{Cascade}}$ times the neutral posterior odds. If we imagine that a court case was decided only on whether the posterior odds on one piece of evidence were greater than 1, we can see that individuals could face higher chances of being found guilty than the evidence would warrant. We will discuss in the following sections how the fact that cases are based on multiple pieces of evidence, each of which is subject to bias, actually aggravates this issue. 

\subsection{Bias snowball}

The \textit{bias snowball effect}, occurs when ``bias increases as irrelevant information from a variety of sources is integrated and influences each other''\citep{drorpyramid}. In this section, we describe how the sorts of errors we have discussed so far can snowball, building on one another and ultimately affecting the final determination of guilt. 

\begin{figure}[ht]
   \centering
   \includegraphics[width=.8\textwidth]{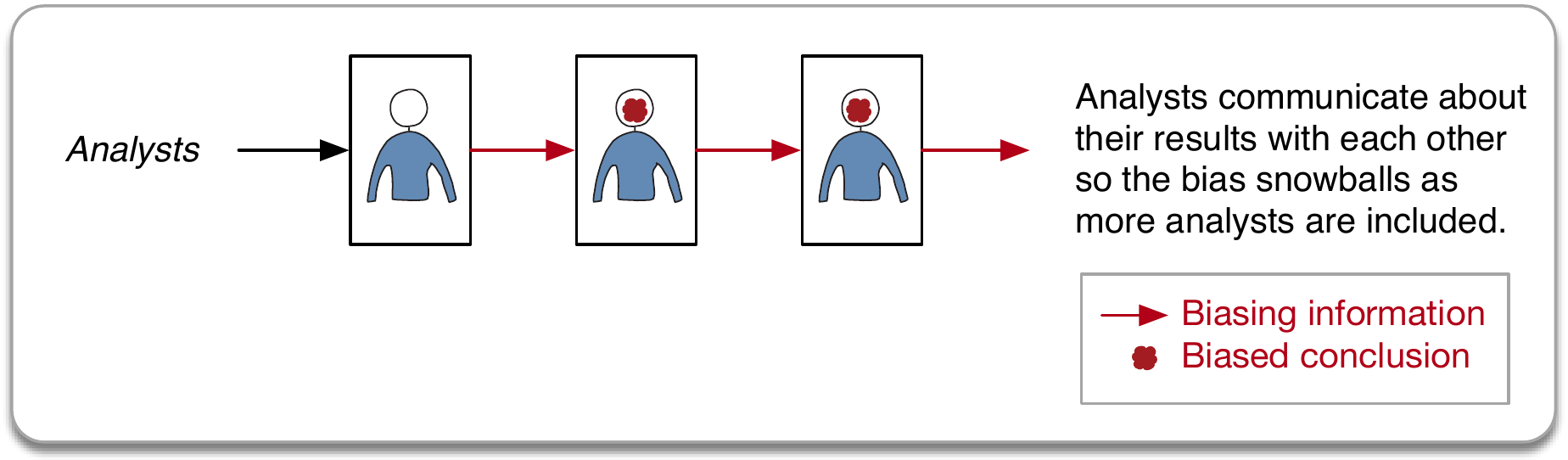} 
   \caption{Bias snowball example. An analyst shares their conclusion with the next analyst, who shares their conclusion with the next. The communication between examiners produces bias, and this accumulates or ``snowballs'' as more analysts are included.}
   \label{fig:biassnowball}
\end{figure}

Fig. \ref{fig:biassnowball} shows that as information is passed from one examiner to the next (either with or without additional task-irrelevant information), and thus the bias grows quickly, like a snowball as it rolls down a hill. The figure only shows communication going in one direction, but in reality examiners could be altering each other's conclusions, and thus the mechanism could be quite complex. Note that if there is a bias cascade, and the examiners communicate with each other, this is equivalent to a bias snowball since the snowball relates to the communication between analysts.

In contrast to bias cascade, bias snowball comes from different lines of evidence informing one another: ``when one piece of forensic evidence (biased or not) is known to other forensic	analysts who are analyzing different forensic evidence, and their examination is affected and biased by their knowledge of the results of the other lines of evidence.'' \citep{dror2017letter} For example, the fingerprint analyst is aware of DNA evidence that points to the guilt of the suspect, and in turn the analyst looking at blood spatter knows the results of the fingerprint analysis. 

Mathematically, this is very similar to the bias cascade setting, but conceptually, it arises from different practices and may have different impacts. In this case, although it is true that the chances of a fingerprint match are higher given a DNA match, using the former to inform determination of the latter double counts this evidence in the final determination of guilt. 

We saw that for one piece of evidence, including true but irrelevant information can lead to a biased estimate of the posterior. Assume the first analyst produces estimated posterior odds, denoted by hat, 
$
\hat{M}_1=\frac{P_A(E_0\mid I_1, X_1, Y_1)}{P_A(E_0^\mathsf{c}\mid I_1, X_1, Y_1)}
$
, and passes these odds on to the second analyst. The neutral posterior odds, which the second analyst cannot see, we denote, 
$
M_1 = \frac{P(E_0\mid I_1, X_1, Y_1)}{P(E_0^\mathsf{c}\mid I_1, X_1, Y_1)}.
$
Now the analyst of the information has access to $\hat{M}_1$ as well as some $I_2$ (which may or may not be the exact same information as what was given to the first analyst). The posterior of the second analyst now has a two-step bias. First, the posterior,
$
\hat{M}_2= \frac{P_{A_2}(E_0 \mid X_2,Y_2,I_2,\hat{M}_1)}{P_{A_2}(E_0^\mathsf{c} \mid X_2,Y_2,I_2,\hat{M}_1)}
$
is a biased estimator of the posterior they would arrive at were they able to see an unbiased posterior, denoted by tilde,
$
    \Tilde{M}_2 = \frac{P_{A_2}(E_0 \mid X_2,Y_2,I_2,M_1)}{P_{A_2}(E_0^\mathsf{c} \mid X_2,Y_2,I_2,M_1)},
$
which just as before is a biased estimator of the posterior,
$
    M_2 = \frac{P(E_0 \mid X_2,Y_2,I_2,M_1)}{P(E_0^\mathsf{c} \mid X_2,Y_2,I_2,M_1)}.
$

The second difference between $M_2$ and $\Tilde{M}_2$ can be expressed just as we have seen earlier by treating $M_1$ as a component of $I_2$. That is, $M_1$ is true but task-irrelevant information (it has no impact on the appearance of the evidence). Thus we can write,
\begin{align*}
    \frac{P_{A_2}(E_0 \mid X_2,Y_2,I_2,M_1)}{P_{A_2}(E_0^\mathsf{c} \mid X_2,Y_2,I_2,M_1)} &= \delta_2(I,M_1) \cdot \frac{P(E_0 \mid X_2,Y_2,I_2,M_1)}{P(E_0^\mathsf{c} \mid X_2,Y_2,I_2,M_1)}\\
    \Rightarrow \Tilde{M}_2 &= \delta_2(I,M_1) M_2.
\end{align*}

To expand on the bias between $\tilde{M}_2$ and $\hat{M}_2$, we can break down the second analyst's final posterior estimation $\hat{M}_2$ as before,
\begin{align}
    \hat{M}_2 &= \frac{P_{A_2}(E_0 \mid x_2,y_2,I_2,\hat{M}_1)}{P_{A_2}(E_0^\mathsf{c} \mid x_2,y_2,I_2,\hat{M}_1)} \\
    &= \frac{P_{A_2}(x_2,y_2,I_2,\hat{M}_1\mid E_0)}{P_{A_2}( x_2,y_2,I_2,\hat{M}_1\mid E_0^\mathsf{c})} \cdot \frac{P_{A_2}(E_0)}{P_{A_2}(E_0^\mathsf{c})}\\
    &=
    \underbrace{\frac{P_{A_2}(x_2,y_2\mid E_0,I_2,\hat{M}_1)}
    {P_{A_2}( x_2,y_2\mid E_0^\mathsf{c},I_2,\hat{M}_1)}}_I
    \cdot
    \underbrace{\frac{P_{A_2}(I_2\mid E_0,\hat{M}_1)}{P_{A_2}(I_2\mid E_0^\mathsf{c},\hat{M}_1 )}}_{II}
    \cdot
    \underbrace{\frac{P_{A_2}(\hat{M}_1\mid E_0)}{P_{A_2}(\hat{M}_1\mid E_0^\mathsf{c})}}_{III}
    \cdot
    \underbrace{\frac{P_{A_2}(E_0)}{P_{A_2}(E_0^\mathsf{c})}}_{IV}.
\end{align}

Term I acts like the likelihood we saw earlier, with the additional conditioning that the first analyst produced a posterior odds of $\hat{M}_1$. We might imagine that the analyst feels even more tempted, perhaps subconsciously, to impute data to match the findings of the first analyst. We'll call this bias term $\Tilde{\delta}_{\text{Impute}}(m)$ and let it depend on the reported posterior from the first analyst. So for example if $\hat{M}_1$ is very high (meaning the first analyst is firmly convinced the evidence points to the suspect), the second analyst may feel comfortable filling in more of the data than they would have with just $I_2$, or with a low value of $\hat{M}_1$. Thus,
$
    \frac{P_{A_2}(x_2,y_2\mid E_0,I_2,\hat{M}_1)}{P_{A_2}( x_2,y_2\mid E_0^\mathsf{c},I_2,\hat{M}_1)} = \Tilde{\delta}_{\text{Impute}}(\hat{M}_1)\frac{P_{A_2}(x_2,y_2\mid E_0,I_2,M_1)}{P_{A_2}( x_2,y_2\mid E_0^\mathsf{c},I_2,M_1)}.
$
In a similar vein, the prior term has a bias of $\Tilde{\delta}_{I}(m)$ where,
$
    \frac{P_{A_2}(I_2\mid E_0,\hat{M}_1)}{P_{A_2}(I_2\mid E_0^\mathsf{c},\hat{M}_1 )} =  \Tilde{\delta}_{I}(\hat{M}_1)\frac{P_{A_2}(I_2\mid E_0,M_1)}{P_{A_2}(I_2\mid E_0^\mathsf{c},M_1)}.
$

Term III is a measure of the second analyst's confidence in the first analyst's abilities. If the second analyst is not correcting for biases may have contaminated the first analyst's output, this ratio will again be biased. We write this as,
$
    \frac{P_{A_2}(\hat{M}_1 \mid E_0)}{P_{A_2}(\hat{M}_1 \mid E_0^\mathsf{c})} = \Tilde{\delta}_{\text{Peer}}(\hat{M}_1) \frac{P_{A_2}(M_1 \mid E_0)}{P_{A_2}(M_1 \mid E_0^\mathsf{c})}.
$
Substituting back in to the full posterior gives,
\begin{align}
    \hat{M}_2 = \Tilde{\delta}_2(\hat{M}_1,I) \Tilde{M}_2,
\end{align}
Where $\Tilde{\delta}_2(\hat{M}_1) = \Tilde{\delta}_{\text{Peer}}\cdot \Tilde{\delta}_{I}(\hat{M}_1) \cdot \Tilde{\delta}_{\text{Impute}}(\hat{M}_1)$. 

To summarize, in the snowball case, bias can arise in two main ways. First, there is the bias we would see even if an analyst were given only true information. This is what we saw in the case of a bias cascade, except that the true information could now include posterior odds about previously analyzed pieces of evidence. Second, there is the bias introduced when an analyst has access to only estimates of true information--namely the judgement of their peers about the evidence. We use an overbar to denote a history, so $\bar{M}_i$ denotes all of the results up to the $i$'th piece of evidence.

Let $$\tilde{\delta}_i(I,x_i,y_i, \bar{\hat{M}}_i) = \Tilde{\delta}_{\text{Impute}}(I,x_i,y_i, \bar{\hat{M}}_i)\Tilde{\delta}_{I}(\bar{\hat{M}}_i)\Tilde{\delta}_{\text{Peer}}(\bar{\hat{M}}_i)$$

\noindent and let 
$$\delta_i(I, x_i,y_i,\bar{M_i}) = \delta_{\text{Impute}}(I,x_i,y_i, \bar{M_i})\delta_{I}(I,\bar{M_i})\delta_{\text{Peer}}(\bar{M_i}).$$
\noindent Then,
\begin{align*}
    \hat{M}_2 &= \Tilde{\delta}_2(\hat{M}_1,x_i,y_i,I) \cdot \Tilde{M}_2 \\
    &= \Tilde{\delta}_2(\hat{M}_1,x_i,y_i,I) \cdot \delta_2(M_1,x_i,y_i,I) \cdot  M_2
\end{align*}

The first term $\Tilde{\delta}_2(\hat{M}_1)$ is the bias from seeing previous analyst's biased conclusions. The second term $\delta_2(M_1,I)$ is the bias from seeing any irrelevant information, even if it is true. As the number of pieces of evidence grows and later analysts are shown more and more information from previous analysts, it is clear these biases will snowball.

\begin{figure}[htb!]
    \centering
    \includegraphics[width = .9\textwidth]{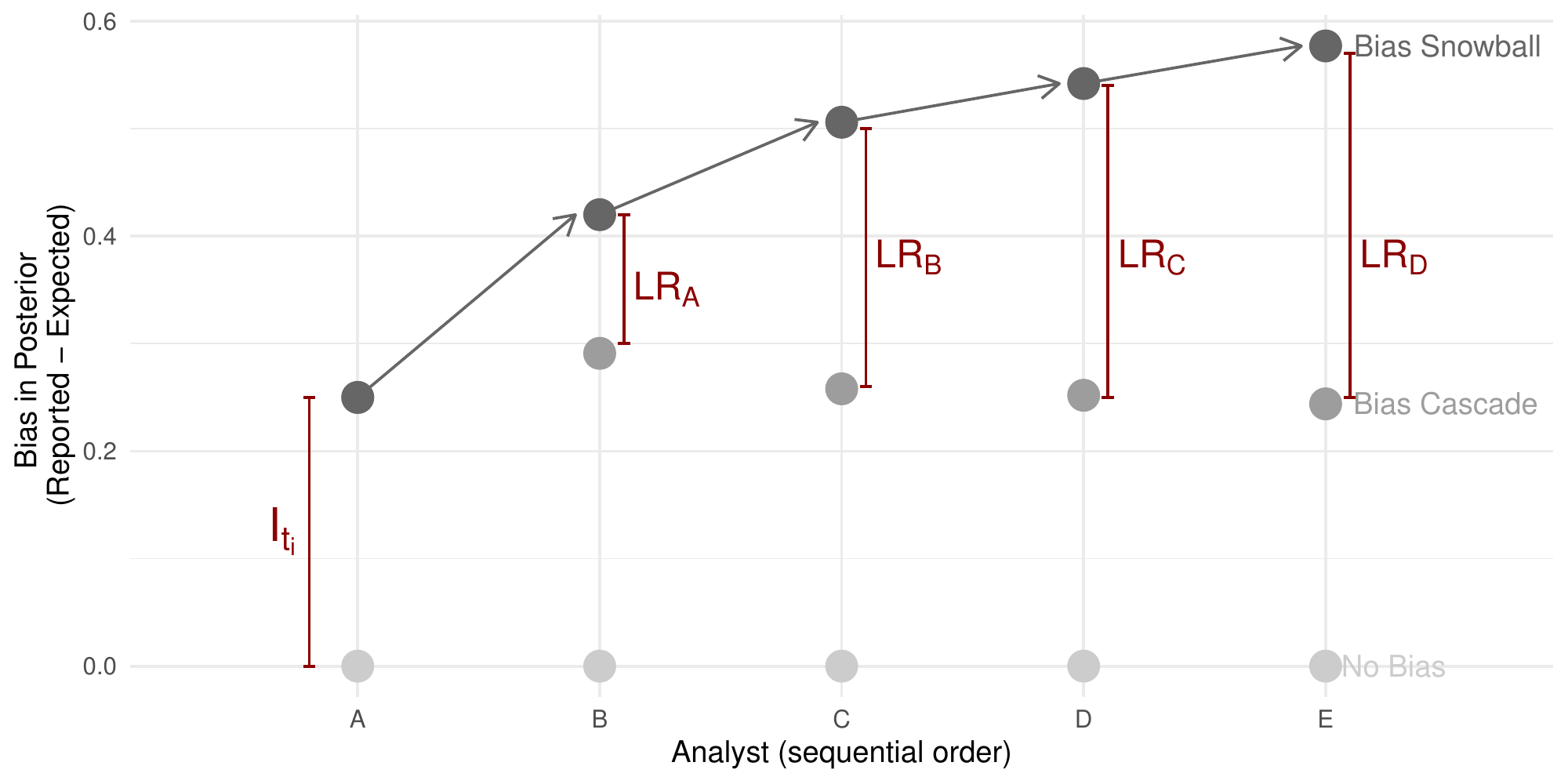}
    \caption{Comparison of bias snowball and bias cascade in a simulation in which five analysts sequentially examine a type of evidence after being exposed to task-irrelevant information ($I_{t_{i}}$). In the bias snowball case, analysts are exposed to the conclusions from the prior analyst ($LR_{X}$, transfer of information illustrated with arrows). Once bias has cascaded or snowballed, it is impossible to return to an unbiased estimate of the posterior.}
    \label{fig:Snowball2}
\end{figure}

We can run another illustrative simulation to show how this might happen. We imagine a scenario where we have $N=10$ suspects, and one characteristic $I$ such that $P(I=1)=0.15$ and as before $P(E_0 \mid I) = P(E_0)$. We will assume in this case that the analysts across $K=5$ pieces of evidence don't have additional bias if they see true information about matches, but true information about $I$ and missing values can still lead to bias. We also assume seeing estimated matches doesn't bias analysts' through imputation or compound their bias with regards to $I$, but that they do over-weight the importance of matches. Let $\mathbb{P}_n R$ denote the share of missing data. The bias terms discussed above are therefore,\footnote{Recall that these are multiplicative biases so a term of 1 means no additional bias}
\begin{align*}
    \delta_{\text{Impute}}(I,x_i,y_i, \bar{M_i}) &=  \mathbb{P}_n R + 0.5*I + 1\\
    \delta_{I}(I,\bar{M_i}) &= 2 - \frac{1-I}{1-P(I=1)}\\
    \delta_{\text{Peer}} &= 1\\
    \Tilde{\delta}_{\text{Impute}}(I,x_i,y_i, \bar{\hat{M}}_i) &= 1\\
    \Tilde{\delta}_{I}(\bar{\hat{M}}_i) &= 1\\
    \Tilde{\delta}_{\text{Peer}}(\bar{\hat{M}}_i) &= \sum_i \indc{\hat{M}_1 \geq 1} + 1.
\end{align*}
We model each of the $k$ pieces of evidence as an independent draw from a Binomial. We assume that if $x$ and $y$ come from the same source, the chance of a match for piece is 0.5; if they come from different sources, it is 0.25. After 1,000 simulations, we see in Fig. \ref{fig:Snowball2} 

\section{Systemic bias: Bias the in final determination of guilt by the trier of fact}\label{sec:finalDet}

Recall that for the trier of fact (i.e., a judge or jury), the final goal is to determine $\frac{P(G\mid \boldsymbol{M})}{P(G^c\mid \boldsymbol{M})}$, where $\boldsymbol{M} = \{ M_1, \cdots, M_k,I \}$ includes all strains of physical evidence as interpreted by the forensic analysts $M_i$ as well as contextual evidence $I$. Once the evidence $\boldsymbol{M}$ gets to the trier of fact, biases that $\boldsymbol{M}$ had from previous processes are "baked in", and it is difficult to know where the biases came from originally. 

Thus, we show that this bias is \textit{systemic}: it arises in various stages throughout the system, and affects the final conclusions. The trier of fact does not know that the experts have biased opinions and takes them each as unbiased estimates. We have used relatively simple probabilistic ideas to show that systemic bias is not only possible but is actually quite likely to occur in this part of the criminal justice system.

\begin{figure}[ht]
   \centering
   \includegraphics[width=.8\textwidth]{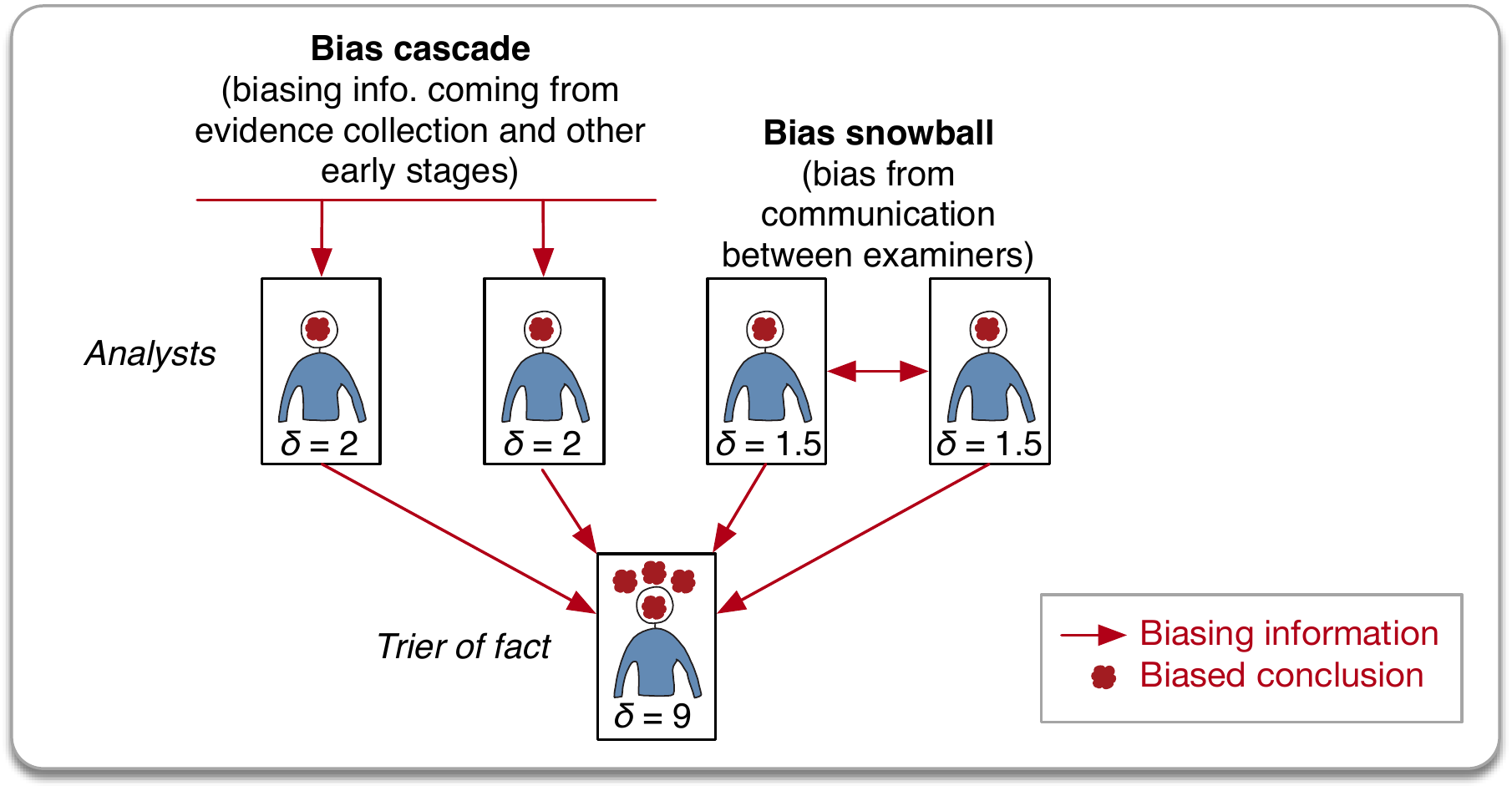} 
   \caption{Bias propagation to the trier of fact. Various forms of bias in the examiners' reports can lead to the trier of fact having an accumulation of bias. Bias can increase as the amount of task-irrelevant information or the number of examiners communicating with each other increase.}
   \label{fig:biasfinal}
\end{figure}

Fig. \ref{fig:biasfinal} shows how the bias is propagated by the examiners and ends up being compounded by the trier of fact, unbeknownst to both examiners and trier of fact. In this example, the examiners were subject to both a bias cascade and a bias snowball. Note that the conclusions from the different types of evidence (e.g., fingerprints, firearms, DNA, shoe prints, etc.) could be weighted more or less heavily depending on how much the trier of fact trusts that the forensic conclusion is correct. For example, a sample of US residents believed that conclusions from DNA are 83\% accurate, fingerprints are 78\%, firearms and toolmarks are 68\%, and voice recognition is 55\% \citep{lingkaplancuellar}. Studies have shown that accuracy, when it has been estimated, is often lower in most disciplines \citep{pcast}. Conclusions such as these could be used to devise a system of weights for how triers of fact could weigh conclusions from different forensic disciplines. Furthermore, different types of triers of fact could weigh evidence differently. For instance, a juror drawn from the general population  might have different beliefs about the validity of a forensic discipline--the CSI effect could be driving this \citep{cole}--than an judge with years of experience hearing forensic testimony. For our purposes, we do not include specific weights for conclusions from different disciplines.

In the final determination of guilt, the posterior odds can be written as
\begin{align}
    \frac{P(G\mid \boldsymbol{M})}{P(G^\mathsf{c} \mid \boldsymbol{M})} 
    &= \frac{P(G\mid \cup_k M_k, I)}{P(G^\mathsf{c} \mid \cup_k M_k, I)}
    \\
    &= \frac{P(\cup_k M_k\mid I, G)}{P(\cup_k M_k\mid I, G^\mathsf{c})}\frac{P(I\mid G)}{P(I\mid G^\mathsf{c})}\frac{G}{G^\mathsf{c}}
    \\ 
    &= \frac{P(\cup_k M_k\mid I, G)}{P(\cup_k M_k\mid I, G^\mathsf{c})}\frac{P(I\mid G)}{P(I\mid G^\mathsf{c})}\frac{1}{N}. 
\end{align}
We assume that each piece of evidence is related to all others only through the guilt (or innocence) of the suspect and contextual evidence. For example, the chances that both the fingerprints match and the DNA matches are only related because the suspect is guilty and perhaps factors about the suspect themselves contained in $I$. This implies a neutral posterior,
\begin{align}\label{eq:neutralGuilt}
    \frac{P(G\mid I,\boldsymbol{M})}{P(G^\mathsf{c} \mid I,\boldsymbol{M})}
    &= \frac{1}{N} \frac{P(M_1\mid I, G)  }{P(M_1\mid I, G^\mathsf{c})  }\frac{P(M_2\mid I, G)}{P(M_2\mid I, G^\mathsf{c})} \cdots \frac{P(M_k\mid I, G)}{P(M_k\mid I, G^\mathsf{c})}\frac{P(I\mid G)}{P(I\mid G^\mathsf{c})}.
\end{align}
The proper way to calculate the chances of guilt is therefore to estimate the probability of each piece of evidence separately. 

Imagine instead that the results of one stream of evidence influence beliefs about the next stream of evidence. We have seen the effects of this ``bias cascade.'' Then we saw the effects when the strains of evidence are allowed to influence one another when discussing ``bias snowball.'' A jury would likely be unaware that the strains of evidence have contaminated one another, and will treat each stream of evidence as if it were independent of the others. With these biases in place, instead of the neutral posterior odds given in (\ref{eq:neutralGuilt}), we have, 
\begin{align}
\label{eq:Guilt}
    \frac{P(G\mid \boldsymbol{\hat{M}})}{P(G^\mathsf{c}\mid \boldsymbol{\hat{M}})} &= \frac{1}{N} \frac{P(\hat{M_1}\mid I, G) }{P(\hat{M_1}\mid I, G^\mathsf{c}) }\frac{P(\hat{M_2}\mid I, G)}{P(\hat{M_2}\mid I, G^\mathsf{c})} \cdots \frac{P(\hat{M_k}\mid I, G)}{P(\hat{M_k}\mid I, G^\mathsf{c})}\frac{P(I\mid G)}{P(I\mid G^\mathsf{c})} \\
    &= \frac{1}{N} \left[ \beta_1(H_1) \frac{P(M_1\mid I, G)}{P(M_1\mid I, G^\mathsf{c})}\right]\cdot \left[\beta_2(H_2)\frac{ P(M_2\mid I, G)}{P(M_2\mid I, G^\mathsf{c})}\right] \cdots \\ 
    &\left[\beta_k(H_{k})\frac{P(M_k\mid I, G)}{P(M_k\mid I, G^\mathsf{c})}\right]\frac{P(I\mid G)}{P(I\mid G^\mathsf{c})}\\
    &=  \prod_{j=1}^k \beta_j(H_{j})\frac{P(G\mid 
    \boldsymbol{M})}{P(G^\mathsf{c} \mid \boldsymbol{M})}.
\end{align}

In this context, the first term $\beta_1(H_1) = \beta_1(I)$ measures the biased chances of seeing a first match declared because $I$ was incorporated into the first analyst's decision-making. The $\beta_j(H_{j})$ terms represent the biased chances of seeing a match declared by the $j$'th analyst due to having been given access to information about all previous analyst's beliefs.

We assume the triers of fact in this setting hold no bias of their own, but that they treat each piece of evidence as if it had been analyzed independently of any other information. For instance, they believe that the DNA analyst did not use the fingerprint analyst's conclusions to inform their own. If this assumption does not hold, then it will only mean that it will be more difficult for the trier of fact to make the correct decision on average.

Thus, we have shown how biases aggregate and interact throughout the investigation and legal process, and end in a larger, compounded bias in the final determination of guilty by the trier of fact: a clear example of systemic bias.

\section{Conclusion}

We have formalized the definitions of bias at the level of the individual examiner and the laboratory demonstrated in previous empirical studies. For an individual examiner, we give probabilistic formalizations of imputation bias and of the long-term effects of bias in an analyst's prior. For the propagation of bias among analysts, we give formalizations of bias cascade and bias snowball. And for the trier of fact, we give a formalization of how biases from individual examiners and groups of examiners compound and lead to systemic bias.

Our formalization is based on years of research, mostly in the field of psychology, and it extends this research by demonstrating how biases can be propagated throughout the system to impact the decisions made by the trier of fact. In this way, we have shown that not only is systemic bias possible, but under certain conditions it necessarily will occur. For instance, if task-irrelevant information has led to bias at the examiner level, this will propagate and possibly compound such that the final determination of guilt is biased.

We have shown that contextual bias in forensics is not only a qualitative problem, and that it can quantitatively lead to a serious compounding of errors. By formalizing contextual bias, we clarify where errors arise and when forensic analysts should be given which information. Without a formal framework for discussing bias, a plethora of types of biases might be used in overlapping or ambiguous ways. Our contribution is to by define biases in rigorous, probabilistic terms to help to eliminate vagueness and ambiguity. Not only can this. We motivate this analysis with the study of fingerprint matching in forensic science, an example in which analysts may be given additional truthful information about suspects which biases their determination of whether the fingerprints are from the same source.

One example in which our research could be used is algorithm design. Algorithms are increasingly used in forensic science to inform and complement examiners' conclusions. For instance, there are a number of Automated Fingerprint Identification Systems (AFIS). The US Integrated AFIS holds the fingerprint sets collected in the United States, and is managed by the FBI. Many states also have their own AFISs. AFISs have capabilities such as latent searching, minutiae identification, electronic image storage, and electronic exchange of fingerprints and responses.

In order for the widespread adoption of algorithms in forensics to be most effective, it is important to be aware of how the errors are occurring in human-based decisions so they can be prevented from happening in algorithms. As researchers develop more of these algorithms, they need to be aware of how human biases enter the final conclusions and initial inputs, and formal tools for discussing and preventing bias are needed. Only by knowing how to formalize the biases observed empirically can algorithms improve upon human performance.

\null \newpage
\bibliographystyle{apa}
\bibliography{foundationspaper.bib}

\begin{thebibliography}{}

\bibitem[\protect\astroncite{Abraham et~al.}{2013}]{abraham2013modern}
Abraham, J., Champod, C., Lennard, C., and Roux, C. (2013).
\newblock Modern statistical models for forensic fingerprint examinations: a
  critical review.
\newblock {\em Forensic Science International}, 232(1-3):131--150.

\bibitem[\protect\astroncite{Aitken}{2018}]{Aitken2018}
Aitken, C. (2018).
\newblock Bayesian hierarchical random effects models in forensic science.
\newblock {\em Frontiers in Genetics}, 9:126.

\bibitem[\protect\astroncite{Cole and Dioso-Villa}{2008}]{cole}
Cole, S. and Dioso-Villa, R. (2008).
\newblock Investigating the csi effect effect: Media and litigation crisis in
  criminal law.
\newblock {\em Stan. L. Rev.}, 61:1335.

\bibitem[\protect\astroncite{Council et~al.}{1996}]{national1996evaluation}
Council, N.~R. et~al. (1996).
\newblock {\em The evaluation of forensic DNA evidence}.
\newblock National Academies Press.

\bibitem[\protect\astroncite{Dror}{2020}]{drorpyramid}
Dror, I.~E. (2020).
\newblock Cognitive and human factors in expert decision making: Six fallacies
  and the eight sources of bias.
\newblock {\em Analytical Chemistry}, 92:7998--8004.

\bibitem[\protect\astroncite{Dror et~al.}{2006}]{dror-mayfield}
Dror, I.~E., Charlton, D., and Peron, A.~E. (2006).
\newblock Contextual information renders experts vulnerable to making erroneous
  identifications.
\newblock {\em Forensic Science International}, 156:74--78.

\bibitem[\protect\astroncite{Dror and Hampikian}{2011}]{dror-dnaevidence}
Dror, I.~E. and Hampikian, G. (2011).
\newblock Subjectivity and bias in forensic {DNA} mixture interpretation.
\newblock {\em Science \& Justice}, 51(4):204--208.

\bibitem[\protect\astroncite{Dror et~al.}{2017}]{dror2017letter}
Dror, I.~E., Morgan, R.~M., Rando, C., and Nakhaeizadeh, S. (2017).
\newblock Letter to the editor--the bias snowball and the bias cascade effects:
  Two distinct biases that may impact forensic decision making.
\newblock {\em Journal of forensic sciences}, 62(3):832--833.

\bibitem[\protect\astroncite{Gross et~al.}{2014}]{kennedy}
Gross, S.~R., O’Brien, B., Hu, C., and Kennedy, E.~H. (2014).
\newblock Rate of false conviction of criminal defendants who are sentenced to
  death.
\newblock {\em Proceedings of the National Academy of Sciences},
  111(20):7230--7235.

\bibitem[\protect\astroncite{Kaye}{2015}]{kayeblog}
Kaye, D. (2015).
\newblock Blinding forensic analysts to task-irrelevant information: A national
  commission (ncfs) speaks out.
\newblock Online.

\bibitem[\protect\astroncite{Lindley}{1977}]{lindley}
Lindley, D.~V. (1977).
\newblock A problem in forensic science.
\newblock {\em Biometrika}, 64(2):207--213.

\bibitem[\protect\astroncite{Ling et~al.}{2020}]{lingkaplancuellar}
Ling, S., Kaplan, J., and Cuellar, M. (2020).
\newblock Public beliefs about the accuracy and importance of forensic evidence
  in the united states.
\newblock {\em Science \& Justice}, 60(3):263--272.

\bibitem[\protect\astroncite{Loeffler et~al.}{2019}]{innocenceestimate}
Loeffler, C.~E., Hyatt, J., and Ridgeway, G. (2019).
\newblock Measuring self-reported wrongful convictions among prisoners.
\newblock {\em Journal of Quantitative Criminology}, 35(2):259--286.

\bibitem[\protect\astroncite{Lund and Iyer}{2016}]{lundiyer}
Lund, S.~P. and Iyer, H. (2016).
\newblock Likelihood ratio as weight of forensic evidence: A metrological
  perspective.
\newblock {\em arXiv}, 1608.07598v2:1--41.

\bibitem[\protect\astroncite{{National Commission on Forensic
  Science}}{2015}]{nationalcommission}
{National Commission on Forensic Science} (2015).
\newblock Ensuring that forensic analysis is based upon task-relevant
  information.
\newblock Views of the Commission.

\bibitem[\protect\astroncite{NRC}{2009}]{nationalacademy}
NRC (2009).
\newblock {\em National Research Council, Strengthening forensic science in the
  United States: a path forward}.
\newblock National Academies Press.

\bibitem[\protect\astroncite{Ommen and Saunders}{2018}]{ommen}
Ommen, D.~M. and Saunders, C.~P. (2018).
\newblock Building a unified statistical framework for the forensic
  identification of source problems.
\newblock {\em Law, Probability and Risk}, 17(2):179--197.

\bibitem[\protect\astroncite{Osborne et~al.}{2014}]{osborne-bitemark}
Osborne, N., Woods, S., Kieser, J., and Zajac, R. (2014).
\newblock Does contextual information bias bitemark comparisons?
\newblock {\em Science \& Justice}, 54(4):267--273.

\bibitem[\protect\astroncite{PCAST}{2016}]{pcast}
PCAST (2016).
\newblock {\em President's Council of Advisors on Science and Technology,
  Report to the President, Forensic Science in Criminal Courts: Ensuring
  Scientific Validity of Feature-comparison Methods (Last accessed July 1,
  2017)}.
\newblock Executive Office of the President of the United States, President's
  Council.

\bibitem[\protect\astroncite{Steele and Balding}{2014}]{steele2014dna}
Steele, C.~D. and Balding, D.~J. (2014).
\newblock Statistical evaluation of forensic dna profile evidence.
\newblock {\em Annual Review of Statistics and Its Application}, 1(1):361--384.

\bibitem[\protect\astroncite{Stockmarr}{1999}]{stockmarr_likelihood_1999}
Stockmarr, A. (1999).
\newblock Likelihood ratios for evaluating {DNA} evidence when the suspect is
  found through a database search.
\newblock {\em Biometrics}, 55(3):671--677.

\bibitem[\protect\astroncite{{The Innocence Project}}{2019}]{innocenceproject1}
{The Innocence Project} (2019).
\newblock Overturning wrongful convictions involving misapplied forensics (last
  accessed: September 30, 2019).
\newblock
  \url{https://www.innocenceproject.org/overturning-wrongful-convictions-involving-flawed-forensics/}.

\bibitem[\protect\astroncite{Thompson et~al.}{2013}]{thompsonbias}
Thompson, W., Vuille, J., Biedrmann, A., and Taroni, F. (2013).
\newblock The role of prior probability in forensic assessments.
\newblock {\em Frontiers in Genetics}, 4:220--223.

\bibitem[\protect\astroncite{{US Department of Justice, Office of the Inspector
  General}}{2006}]{fbi}
{US Department of Justice, Office of the Inspector General} (2006).
\newblock A review of the fbi's handling of the brandon mayfield case.
\newblock Unclassified Executive Summary.

\bibitem[\protect\astroncite{Willis et~al.}{2015}]{ensfireport}
Willis, S., McKenna, L., Mc~Dermott, S., O'Donnell, G., Barrett, A.,
  Rasumusson, B., Hoglund, T., Nordgaard, A., Berger, C., Sjerps, M.,
  Lucena~Molina, J., Zadora, G., Aitken, C., Lovelock, T., Lunt, L., Champod,
  C., Bidermann, A., Hicks, T., and Taroni, F. (2015).
\newblock {ENFSI} guideline for evaluative reporting in forensic science.

\end{thebibliography}

\end{document}